# Completely-positive quantum operations generating thermostatistical states: A comparative study

Yasuyuki Matsuo, Sumiyoshi Abe

*Department of Physical Engineering*, *Mie University*, *Mie 514-8507*, *Japan*

**Abstract**   Nonunitary quantum operations generating thermostatistical states and forming positive operator-valued measures (POVMs) are of current interest as a useful tool for operational approach to quantum thermodynamics. Here, two different operations generating the same thermostatistical state are studied: one is related to thermofield dynamics and the other is the one proposed in a recent work [S. Abe, A. R. Usha Devi, A. K. Rajagopal, J. Phys. A: Math. Theor. 43 (2010) 045303]. A comparable study on them shows different behaviors of the von Neumann entropy under repeated applications of these two operations. It is shown that the entropy does not behave monotonically, in general, and can even decrease under the thermofield-dynamical one, in contrast to monotonic increase under the other.

# 1. Introduction

It is traditional in studies of quantum open systems to start with the isolated composite total system governed by unitary dynamics, divide it into the objective and environmental systems interacting with each other and then eliminate the environment in order to formulate nonunitary subdynamics of the objective system. An important statement in quantum thermodynamics [1], which is currently attracting much attention, is that such a subsystem approaches equilibrium under a wide class of initial conditions. It is however known [2, 3] that the reduced density matrix of the objective system, which is obtained from the total density matrix through the partial trace over the environmental degrees of freedom, may violate positive semidefiniteness, in general, leading to a serious difficulty with the probabilistic interpretation in quantum theory. On the other hand, the heat bath in classical thermodynamics does not play any dynamical role: it is treated rather in an operational manner such as heat exchange with the objective system, i.e., heating and cooling. (In this sense, it is very remarkable even today that the highly-universal laws of classical thermodynamics have been established totally without recourse to dynamics of the bath.) Therefore, it is also physically natural to develop operational approach to quantum thermodynamics. This idea forms a core of a recent work in Ref. [4]. There, an effect of the (hot) heat bath is represented by a heat-up operation given in terms of a completely-positive map proposed in Refs. [5, 6]. A simple thermodynamic process is generated by repeated applications of the operation. In other words, the objective system undergoes a sequence of "discrete measurements" by the environment as an "observer". It has been indicated [4] that the Clausius inequality could be violated along such a process, exhibiting a remarkable quantum effect on thermodynamics.

Now, given a pure state at vanishing temperature, mapping from it to a target mixed state is not unique. This nonuniqueness is a remnant of differences in underlying unitary operations on the total isolated system and may tell us about their thermodynamic implications.

Here, we discuss two different kinds of completely-positive quantum operations, both of which form positive operator-valued measures (POVMs) [7, 8] and generate the same thermostatistical state. One is the statistical quantum operation discussed in Refs. [4-6], and the other is the one derived from so-called thermofield dynamics [9, 10]. To compare them qualitatively and quantitatively, we analyze the behaviors of the von Neumann entropy under repeated applications of them to the pure ground state. We shall see that the entropy exhibits a nonmonotonic behavior and may even decrease under repeated applications of the operation associated with thermofield dynamics, in contrast to monotonic increase in the wide sense under the one in Refs. [4-6]. The result can be interpreted in terms of the concept of unitalness.

The present paper is organized as follows. In Section 2, the unital statistical quantum operation proposed in Ref. [5] is discussed and some explicit forms of its underlying unitary transformations are presented. In Section 3, the operation associated with thermofield dynamics is derived. In section 4, the von Neumann entropy is evaluated for repeated applications of these two different operations. Section 5 is devoted to conclusion. Throughout this paper, both $\hbar$ and $k_B$, the Boltzmann constant, are set equal to unity.

**2. Unital statistical quantum operation and its underlying unitary transformation**

In this section, first we succinctly summarize the operation originally proposed in Ref. [5] and further studied in Refs. [4, 6]. Then, we present some examples of underlying unitary transformations.

Consider a quantum system, $A$, with a Hamiltonian, $H_A$, in $d$ dimensions. The set of the normalized eigenstates, $\{|u_n\rangle_A\}_{n=0,1,...,d-1}$ satisfying $H_A|u_n\rangle_A = \varepsilon_n|u_n\rangle_A$ with the energy eigenvalue $\varepsilon_n$, is assumed to form a complete orthonormal system: $I_A = \sum_{n=0}^{d-1} |u_n\rangle_A {}_A\langle u_n|$, where $I_A$ is the $d \times d$ identity matrix. A quantum operation, $\Phi$, is a map from one density matrix, $\rho_A$, to another: $\rho_A \to \Phi(\rho_A)$, which should be linear, completely positive and trace-preserving. Then, the most general form of $\Phi$ reads

$$\Phi(\rho_A) = \sum_{n=0}^{d-1} V_n \rho_A V_n^\dagger, \qquad (1)$$

where $V_n$'s obey

$$\sum_{n=0}^{d-1} V_n^\dagger V_n = I_A, \qquad (2)$$

which ensures the trace-preserving condition: $\mathrm{Tr}_A \Phi(\rho_A) = \mathrm{Tr}_A \rho_A (\equiv 1)$. $\{V_n^\dagger V_n\}_{n=0,1,...,d-1}$ forms a positive operator-valued measure (POVM).

The operator $V_n$ proposed in Ref. [5] is of the following form:

$$V_n = \sqrt{p_n}\left(I_A - |u_0\rangle_{AA}\langle u_0| - |u_n\rangle_{AA}\langle u_n| + |u_0\rangle_{AA}\langle u_n| + |u_n\rangle_{AA}\langle u_0|\right)$$

$$(n = 0, 1, ..., d-1), \qquad (3)$$

where $p_n$'s satisfy the conditions, $p_n \in (0,1)$ and $\sum_{n=0}^{d-1} p_n = 1$, which guarantee that Eq. (2) in fact holds. The basic structure of this operator is that two terms, $|u_0\rangle_{AA}\langle u_0|$ and $|u_n\rangle_{AA}\langle u_n|$, are picked up from the identity matrix $I_A = \sum_{n'=0}^{d-1} |u_{n'}\rangle_{AA}\langle u_{n'}|$ and are replaced by the transition matrices, $|u_0\rangle_{AA}\langle u_n|$ and $|u_n\rangle_{AA}\langle u_0|$.

$\Phi$ in Eq. (1) with Eq. (3) maps from the pure ground state, $|u_0\rangle_{AA}\langle u_0|$, to the following mixed state:

$$\Phi\left(|u_0\rangle_{AA}\langle u_0|\right) = \sum_{n=0}^{d-1} p_n |u_n\rangle_{AA}\langle u_n|, \qquad (4)$$

in which no off-diagonal elements exist and therefore perfect decoherence is realized. If the form $p_n = \exp(-\beta \varepsilon_n)/Z(\beta)$ with the partition function $Z(\beta) = \sum_{n=0}^{d-1} \exp(-\beta \varepsilon_n)$ is employed, then Eq. (4) becomes the canonical density matrix with the inverse temperature $\beta$: $\exp(-\beta H_A)/Z(\beta)$.

It is shown in Ref. [4] that repeated applications of $\Phi$ to the state in Eq. (4) keep the perfect decoherence unchanged (see Section 4). Furthermore, as the number of repetitions increases, the state approaches the completely-random state realized in the high-temperature limit:

$$\Phi^N\left(|u_0\rangle_{AA}\langle u_0|\right) \to \frac{1}{d} I_A \quad (N \to \infty). \tag{5}$$

This feature is understood as follows. Since $V_n$ in Eq. (3) is Hermitian, it immediately follows from Eq. (2) that

$$\sum_{n=0}^{d-1} V_n V_n^\dagger = I_A \tag{6}$$

also holds. An operation satisfying Eq. (6) is called unital. This implies that the identity matrix is a fixed point:

$$\Phi(I_A) = I_A, \tag{7}$$

which explains Eq. (5). Let $f$ be operator concave, i.e., for any Hermitian matrices, $X$ and $Y$, it satisfies the inequality: $f(\lambda X + (1-\lambda)Y) \geq \lambda f(X) + (1-\lambda)f(Y)$, where $\lambda \in [0,1]$. (A matrix inequality, $X \geq Y$, means that all eigenvalues of $X - Y$ are positive semidefinite.) If $\Phi$ is unital, then the following matrix inequality holds [11, 12]:

$$f(\Phi(X)) \geq \Phi(f(X)). \tag{8}$$

Let $X$ be an arbitrary density matrix $\rho_A$ and consider

$$f(\rho_A) = -\rho_A \ln \rho_A, \tag{9}$$

which is strictly operator concave. Taking the trace of Eq. (8) with Eq. (9), we have

$$S[\Phi(\rho_A)] \geq S[\rho_A], \tag{10}$$

where $S[\rho_A]$ is the von Neumann entropy of the state $\rho_A$ defined by

$$S[\rho_A] = -\text{Tr}(\rho_A \ln \rho_A). \tag{11}$$

The maximum value of the entropy, $S_{\max} = \ln d$, is uniquely reached by the completely-random state realized in the high-temperature limit. This is the reason behind Eq. (5). Thus, repeated applications of $\Phi$ monotonically "heat-up" the system in the pure ground state at vanishing temperature to the state at infinite temperature with intermediate states being nonequilibrium, in general.

Equation (1) is often called the Kraus representation. In the discussion in Ref. [13], the "initial" total density matrix is of the form: $\rho = \rho_A \otimes |\psi\rangle_{EE}\langle\psi|$, where $|\psi\rangle_{EE}\langle\psi|$ is a pure-state density matrix of the environmental system, $E$. Next, this state is unitary-transformed as $\rho \to U\rho U^\dagger$. Then, the partial trace over the environmental degrees of freedom yields the form in Eq. (1) with $V_n$ being given by $V_n = {}_E\langle v_n|U|\psi\rangle_E$, where $\{|v_n\rangle_E\}_n$ is a certain basis in the Hilbert space of $E$. Since the dimensionality of $E$ is the same as that of $A$, it is convenient and sufficient to take as the environment a copy of $A$, as seen in Eq. (3) in the present case. Such a replica is denoted by $B$. Our purpose is to find a unitary matrix, $U$, that can realize the operators in Eq. (3). We write $U$ as follows:

$$U = \sum_{\alpha,\beta=1}^{d^2} u_{\alpha\beta} |\phi_\alpha\rangle\langle\phi_\beta|, \qquad (12)$$

where $\{|\phi_\alpha\rangle\}_{\alpha=1,2,\ldots,d^2}$ is a certain basis of the total Hilbert space of the composite system $(A, B)$ and the unitarity condition leads to

$$\sum_{\gamma=1}^{d^2} u_{\alpha\gamma} u^*_{\beta\gamma} = \delta_{\alpha\beta}.$$

(13)

Also, we employ as the "initial" state the pure ground state at vanishing temperature

$$\rho = |u_0\rangle_{AA}\langle u_0| \otimes |u_0\rangle_{BB}\langle u_0|. \qquad (14)$$

Clearly, $U$ is not unique, in general, and unfortunately it is hard to find even its one explicit form for an arbitrary value of the dimensionality. Therefore, below we only present explicit examples for $d = 2$ and $d = 3$.

In the case $d = 2$, we take the following four unentangled basis states: $|\phi_1\rangle = |u_0\rangle_A |u_0\rangle_B$, $|\phi_2\rangle = |u_0\rangle_A |u_1\rangle_B$, $|\phi_3\rangle = |u_1\rangle_A |u_0\rangle_B$, $|\phi_4\rangle = |u_1\rangle_A |u_1\rangle_B$. Then, an explicit example is

$$(u_{\alpha\beta}) = \begin{pmatrix} \sqrt{p_0} & 0 & 0 & \sqrt{p_1} \\ 0 & -\sqrt{p_0} & \sqrt{p_1} & 0 \\ 0 & \sqrt{p_1} & \sqrt{p_0} & 0 \\ \sqrt{p_1} & 0 & 0 & -\sqrt{p_0} \end{pmatrix}, \tag{15}$$

where $p_0, p_1 \in (0,1)$ and $p_0 + p_1 = 1$. One sees that $V_n = {}_B\langle u_n|U|u_0\rangle_B$ ($n = 0, 1$) yields the result obtained from Eq. (3):

$$V_0 = \sqrt{p_0}\, I_A, \qquad V_1 = \sqrt{p_1}\left(|u_0\rangle_{AA}\langle u_1| + |u_1\rangle_{AA}\langle u_0|\right). \tag{16}$$

In the case $d = 3$, we take the following nine unentangled basis states:

$|\phi_1\rangle = |u_0\rangle_A |u_0\rangle_B$, $\quad |\phi_2\rangle = |u_1\rangle_A |u_0\rangle_B$, $\quad |\phi_3\rangle = |u_0\rangle_A |u_1\rangle_B$, $\quad |\phi_4\rangle = |u_2\rangle_A |u_0\rangle_B$,

$|\phi_5\rangle = |u_1\rangle_A |u_1\rangle_B$, $\quad |\phi_6\rangle = |u_0\rangle_A |u_2\rangle_B$, $\quad |\phi_7\rangle = |u_2\rangle_A |u_1\rangle_B$, $\quad |\phi_8\rangle = |u_1\rangle_A |u_2\rangle_B$,

$|\phi_9\rangle = |u_2\rangle_A |u_2\rangle_B$. A possible solution is given as follows:

$u_{11} = \sqrt{p_0}$, $u_{19} = \sqrt{p_1 + p_2}$, $u_{22} = \sqrt{p_0}$, $u_{23} = \sqrt{p_1}$, $u_{26} = \sqrt{\dfrac{p_0 p_2}{p_0 + p_1}}$,

$u_{27} = \sqrt{\dfrac{p_1 p_2}{p_0 + p_1}}$, $u_{32} = \sqrt{p_1}$, $u_{33} = -\sqrt{p_0}$, $u_{36} = \sqrt{\dfrac{p_1 p_2}{p_0 + p_1}}$, $u_{37} = -\sqrt{\dfrac{p_0 p_2}{p_0 + p_1}}$,

$u_{44} = \sqrt{p_0}$, $u_{48} = -\sqrt{p_1 + p_2}$, $u_{51} = \sqrt{p_1}$, $u_{55} = \sqrt{\dfrac{p_2}{p_1 + p_2}}$, $u_{59} = -\sqrt{\dfrac{p_0 p_1}{p_1 + p_2}}$,

$u_{63} = -\sqrt{\dfrac{p_1 p_2}{p_1 + p_2}}$, $u_{64} = \sqrt{p_2}$, $u_{67} = \sqrt{\dfrac{p_1(p_0 + p_1)}{p_1 + p_2}}$, $u_{68} = \sqrt{\dfrac{p_0 p_2}{p_1 + p_2}}$,

$u_{73} = \dfrac{p_2}{\sqrt{p_1 + p_2}}$, $u_{74} = \sqrt{p_1}$, $u_{77} = -\sqrt{\dfrac{p_2(p_0 + p_1)}{p_1 + p_2}}$, $u_{78} = \sqrt{\dfrac{p_0 p_1}{p_1 + p_2}}$,

$$u_{82} = \sqrt{p_2}, \quad u_{86} = -\sqrt{p_0 + p_1}, \quad u_{91} = \sqrt{p_2}, \quad u_{95} = -\sqrt{\frac{p_1}{p_1 + p_2}}, \quad u_{99} = -\sqrt{\frac{p_0 p_2}{p_1 + p_2}},$$

$$(\text{others}) = 0, \tag{17}$$

where $p_0, p_1, p_2 \in (0,1)$ and $p_0 + p_1 + p_2 = 1$. One can find that $V_n = {}_B\langle u_n | U | u_0 \rangle_B$ ($n = 0, 1, 2$) correctly yields the result

$$V_0 = \sqrt{p_0}\, I_A, \quad V_1 = \sqrt{p_1}\left(|u_2\rangle_{AA}\langle u_2| + |u_0\rangle_{AA}\langle u_1| + |u_1\rangle_{AA}\langle u_0|\right),$$

$$V_2 = \sqrt{p_2}\left(|u_1\rangle_{AA}\langle u_1| + |u_0\rangle_{AA}\langle u_2| + |u_2\rangle_{AA}\langle u_0|\right). \tag{18}$$

Closing this section, we wish to correct a statement made in Ref. [4]. Unlike what is claimed there, the above results manifestly show that the operators in Eq. (3) can be constructed from an unentangled "initial" state in Eq. (14).

## 3. Statistical quantum operation associated with thermofield dynamics

Thermofield dynamics [9, 10] (see Ref. [14] for a more recent work) is also known to generate a thermal state from the pure ground state at vanishing temperature. In this section, we discuss the unitary transformation peculiar in thermofield dynamics, which gives another statistical quantum operation.

Let us consider the harmonic oscillator (with the frequency $\omega$), which is characterized by the creation and annihilation operators, $a^\dagger$ and $a$, satisfying the algebra, $[a, a^\dagger] = I$, $[a, a] = [a^\dagger, a^\dagger] = 0$, and the ground state $|0\rangle$ annihilated by $a$: $a|0\rangle = 0$. In the traditional notation in thermofield dynamics, the replica operators are denoted by $\tilde{a}^\dagger$ and $\tilde{a}$, and the ground state by $|\tilde{0}\rangle$, satisfying the relations isomorphic to the above ones. The original operators and the "tildian" operators commute each other. The thermal state $|O(\theta)\rangle$ is defined by the unitary transformation of $|0\rangle|\tilde{0}\rangle$ as follows:

$$|O(\beta)\rangle = U(\beta)|0\rangle|\tilde{0}\rangle, \qquad (19)$$

$$U(\beta) = \exp\left[\theta(\beta)\left(a^\dagger \tilde{a}^\dagger - a\tilde{a}\right)\right]. \qquad (20)$$

$U(\beta)$ defines the Bogoliubov transformation between the original and tildian operators [9, 10]. The canonical density matrix of the original oscillator is obtained by the partial trace over the tildian degrees of freedom: $\text{Tr}_\sim\left[|O(\beta)\rangle\langle O(\beta)|\right] = \exp(-\beta H)/Z(\beta)$, where $H = \omega\left(a^\dagger a + 1/2\right)$ is the original oscillator Hamiltonian and $Z(\beta) = \text{Tr}\exp(-\beta H) = 1/\left[2\sinh(\beta\omega/2)\right]$ the partition function, provided that $\theta(\beta)$ in Eq. (20) is given by

$$\cosh\theta(\beta) = \frac{1}{\sqrt{1 - e^{-\beta\omega}}}. \qquad (21)$$

Our purpose is to express the canonical density matrix in the form

$$\exp(-\beta H)/Z(\beta) = \sum_{n=0}^{\infty} K_n |0\rangle\langle 0| K_n^\dagger, \qquad (22)$$

where $K_n = \langle \tilde{n} | U(\theta) | \tilde{0} \rangle$ with $|\tilde{n}\rangle$ being the tildian number state defined by $|\tilde{n}\rangle = (\tilde{a}^\dagger)^n |\tilde{0}\rangle / \sqrt{n!}$. To calculate $K_n$, it is convenient to decompose Eq. (20) as follows:

$$\exp\left[\theta(\beta)\left(a^\dagger \tilde{a}^\dagger - a\tilde{a}\right)\right]$$
$$= \exp\left[a^\dagger \tilde{a}^\dagger \tanh\theta(\beta)\right] \exp\left\{-\left(a^\dagger a + \tilde{a}^\dagger \tilde{a} + 1\right) \ln\left[\cosh\theta(\beta)\right]\right\}$$
$$\times \exp\left[-a\tilde{a} \tanh\theta(\beta)\right]. \qquad (23)$$

Then, it is straightforward to obtain

$$K_n = \frac{\left[a^\dagger \tanh\theta(\beta)\right]^n}{\sqrt{n!}\, \cosh\theta(\beta)} \exp\left\{-a^\dagger a \ln\left[\cosh\theta(\beta)\right]\right\} \quad (n = 0, 1, 2, \ldots). \qquad (24)$$

These operators satisfy the trace-preserving condition: $\sum_{n=0}^{\infty} K_n^\dagger K_n = I$, and therefore $\left\{K_n^\dagger K_n\right\}_{n=0,1,2,\ldots}$ forms a POVM. However, the corresponding quantum operation is nonunital, since

$$\sum_{n=0}^{\infty} K_n K_n^\dagger = \frac{1}{\cosh^2\theta(\beta)} I, \qquad (25)$$

which is not equal to the identity matrix.

It is also possible to construct the thermofield-dynamical operation for a system in $d$ dimensions, as in the preceding section. Let us take the following two states:

$$|\Omega\rangle = |u_0\rangle|\tilde{u}_0\rangle, \qquad |\Psi\rangle = \frac{1}{\sqrt{1-p_0}} \sum_{n\neq 0}^{d-1} \sqrt{p_n} \, |u_n\rangle|\tilde{u}_n\rangle, \qquad (26)$$

where $p_n \in (0,1)$ and $\sum_{n=0}^{d-1} p_n = 1$. These states are normalized and are orthogonal to each other. The unitary matrix to be considered is [15, 16]

$$U(\theta) = e^{\theta G}, \qquad (27)$$

where $G$ is an anti-Hermitian matrix given by

$$G = |\Psi\rangle\langle\Omega| - |\Omega\rangle\langle\Psi|. \qquad (28)$$

Using the relations, $G^2 = -\left(|\Omega\rangle\langle\Omega| + |\Psi\rangle\langle\Psi|\right) \equiv -R$ and $G^3 = -GR = -G$, we have

$$U(\theta) = I + G\sin\theta + R(\cos\theta - 1). \qquad (29)$$

With this form, the choice

$$\cos^2\theta = p_0 \qquad (30)$$

leads to

$$\Lambda(|u_0\rangle\langle u_0|) \equiv \text{Tr}_{\sim}\left[U(\theta)|\Omega\rangle\langle\Omega|U^{\dagger}(\theta)\right] = \sum_{n=0}^{d-1} p_n |u_n\rangle\langle u_n|. \tag{31}$$

Thus, perfect decoherence is realized, and the canonical density matrix is obtained if $p_n$ is taken to be the one given after Eq. (4). The corresponding statistical quantum operation reads

$$\Lambda(|u_0\rangle\langle u_0|) = \sum_{n=0}^{d-1} M_n |u_0\rangle\langle u_0| M_n^{\dagger}, \tag{32}$$

where the operator, $M_n = \langle \tilde{u}_n | U(\theta) | \tilde{u}_0 \rangle$, is found to be given by

$$M_n = (I - |u_0\rangle\langle u_0|)\delta_{n,0} + \sqrt{p_n}\, |u_n\rangle\langle u_0|. \tag{33}$$

As in the case of the harmonic oscillator, $\{M_n^{\dagger} M_n\}_{n=0,1,\ldots,d-1}$ also forms a POVM. However, the corresponding operation is not unital, since

$$\sum_{n=0}^{d-1} M_n M_n^{\dagger} = I + \sum_{n=0}^{d-1} p_n |u_n\rangle\langle u_n| - |u_0\rangle\langle u_0|, \tag{34}$$

which is unequal to the identity matrix.

Thus, we see that the statistical quantum operation derived from thermofield dynamics differs from the one discussed in Section 2. In particular, they do not possess the unital property. In the next section, we develop a comparative study on these two different types.

## 4. Behavior of the von Neumann entropy

As mentioned in Section 2, the von Neumann entropy does not decrease under repeated applications of a unital operation. On the other hand, the operation derived from thermofield dynamics is nonunital and therefore expected to yield, in general, a nonmonotonic behavior of the entropy under its repeated applications. This is the issue we are going to discuss in this section.

To notationally distinguish $\Phi$ in Eq. (1) with Eq. (3) and $\Lambda$ in Eq. (32) with Eq. (33), here let us write them anew as follows:

$$\Phi\left(|u_0\rangle\langle u_0|\right) = \sum_{n=0}^{d-1} p_n^{(1)} |u_n\rangle\langle u_n|, \qquad (35)$$

$$\Lambda\left(|u_0\rangle\langle u_0|\right) = \sum_{n=0}^{d-1} q_n^{(1)} |u_n\rangle\langle u_n|, \qquad (36)$$

where $p_n^{(1)}$, $q_n^{(1)} \in (0,1)$ and $\sum_{n=0}^{d-1} p_n^{(1)} = \sum_{n=0}^{d-1} q_n^{(1)} = 1$. It can be shown that, after operating $N$ times, one has the forms

$$\Phi^N\left(|u_0\rangle\langle u_0|\right) = \sum_{n=0}^{d-1} p_n^{(N)} |u_n\rangle\langle u_n|, \qquad (37)$$

$$\Lambda^N\left(|u_0\rangle\langle u_0|\right) = \sum_{n=0}^{d-1} q_n^{(N)} |u_n\rangle\langle u_n|, \qquad (38)$$

where $p_n^{(N)}$, $q_n^{(N)} \in (0,1)$ and $\sum_{n=0}^{d-1} p_n^{(N)} = \sum_{n=0}^{d-1} q_n^{(N)} = 1$. These equations are important, since they imply that the perfect decoherence is kept unchanged under the repeated applications of both $\Phi$ and $\Lambda$ (i.e., eternal absence of off-diagonal terms). Also, one finds that the following recurrence relations hold:

$$p_0^{(N)} = \sum_{n=0}^{d-1} p_n^{(1)} p_n^{(N-1)}, \quad p_k^{(N)} = p_k^{(1)} p_0^{(N-1)} + \left(1 - p_k^{(1)}\right) p_k^{(N-1)} \quad (k = 1, 2, ..., d-1), \quad (39)$$

$$q_n^{(N)} = q_n^{(N-1)} + \left(q_n^{(1)} - \delta_{n,0}\right) q_0^{(N-1)} \quad (n = 0, 1, ..., d-1). \quad (40)$$

Eq. (40) can be explicitly solved as

$$q_0^{(N)} = \left(q_0^{(1)}\right)^N, \quad q_k^{(N)} = \frac{1 - \left(q_0^{(1)}\right)^N}{1 - q_0^{(1)}} q_k^{(1)} \quad (k = 1, 2, ..., d-1). \quad (41)$$

Unfortunately, Eq. (39) is unlikely to be solvable. However, its fixed-point solution can be found and is given by the equiprobability distribution

$$p_n^{(\infty)} = \frac{1}{d} \quad (n = 0, 1, ..., d-1), \quad (42)$$

which corresponds to the completely-random state realized in the high-temperature limit, in accordance with Eq. (7).

To quantitatively compare the physical properties of $\Phi$ and $\Lambda$, we study the behaviors of the von Neumann entropy under their repeated applications:

$$S_N^\Phi \equiv S\left[\Phi^N\left(|u_0\rangle\langle u_0|\right)\right], \tag{43}$$

$$S_N^\Lambda \equiv S\left[\Lambda^N\left(|u_0\rangle\langle u_0|\right)\right]. \tag{44}$$

We have evaluated the values of these quantities by employing a simple three-level system. in particular, we have examined two different "initial" distributions: (i) $p_0^{(1)} = q_0^{(1)} = 2/3$, $p_1^{(1)} = q_1^{(1)} = 1/5$, $p_2^{(1)} = q_2^{(1)} = 2/15$, (ii) $p_n^{(1)} = q_n^{(1)} = 1/3$ ($n = 0, 1, 2$). In Fig. 1, we present the plots of $S_N^\Phi$ with respect to $N$. As discussed in Section 2, the unital nature of $\Phi$ makes the entropy nondecreasing. Under the condition (ii), $S_N^\Phi$ remains constant taking the maximum value, $\ln 3$, since (ii) corresponds to the completely-random state realized in the high-temperature limit, as already mentioned repeatedly. In Fig. 2, we present the plots of $S_N^\Lambda$ with respect to $N$. A nonmonotonic behavior is observed for (i) due to the nonunital nature of $\Lambda$. In particular, $S_N^\Lambda$ has a peak. For both (i) and (ii), $S_N^\Lambda$ converges to a single value. This value is given by Eq. (41) in the limit $N \to \infty$: $p_0^{(N)} \to 0$, $p_k^{(N)} \to p_k^{(1)}/(1-p_0^{(1)})$. Note an exotic feature that the contribution from the ground state tends to disappear. Thus, the limiting value of $S_N^\Lambda$ is of a strongly nonequilibrium state, in contrast to $S_N^\Phi$. In other words, repeated applications of $\Phi$ cover from the vanishing temperature to the high-temperature limit, whereas those of $\Lambda$ cannot cover such a whole range.

## 5. Conclusion

Nonunitary quantum operation generating a desired mixed state from a given pure state is not unique. Here, we have discussed two different kinds of operations, both of which can generate the same thermostatistical state from the pure ground state at vanishing temperature. One, denoted by $\Phi$, possesses the unital nature, whereas the other, $\Lambda$, derived from thermofield dynamics does not, although both of them are completely positive and form POVMs. A point is that the physical difference between them becomes manifest if their repeated applications are considered. To see it quantitatively, we have studied the behavior of the von Neumann entropy under repeated applications of these operations. We have shown that the entropy does not behave monotonically under repeated applications of $\Lambda$, in general, and can even decrease, whereas $\Phi$ makes the entropy nondecreasing because of its unital nature. The results may be useful if these operations are employed as tools for developing operational approach to quantum thermodynamics.

**Acknowledgment**

S. A. was supported in part by a Grant-in-Aid for Scientific Research from the Japan Society for the Promotion of Science.

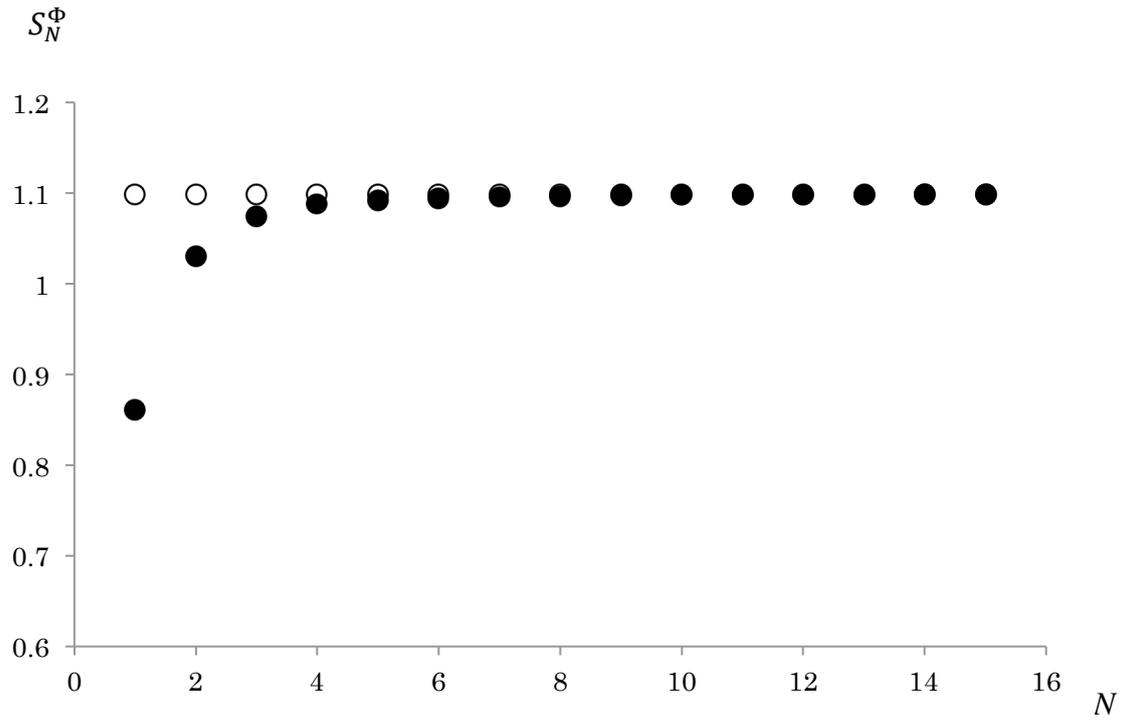

**Fig. 1.** Plots of $S_N^\Phi$ in Eq. (43) with respect to $N$. ● and ○ correspond to the distributions (i) and (ii) mentioned in the text, respectively. All quantities are dimensionless.

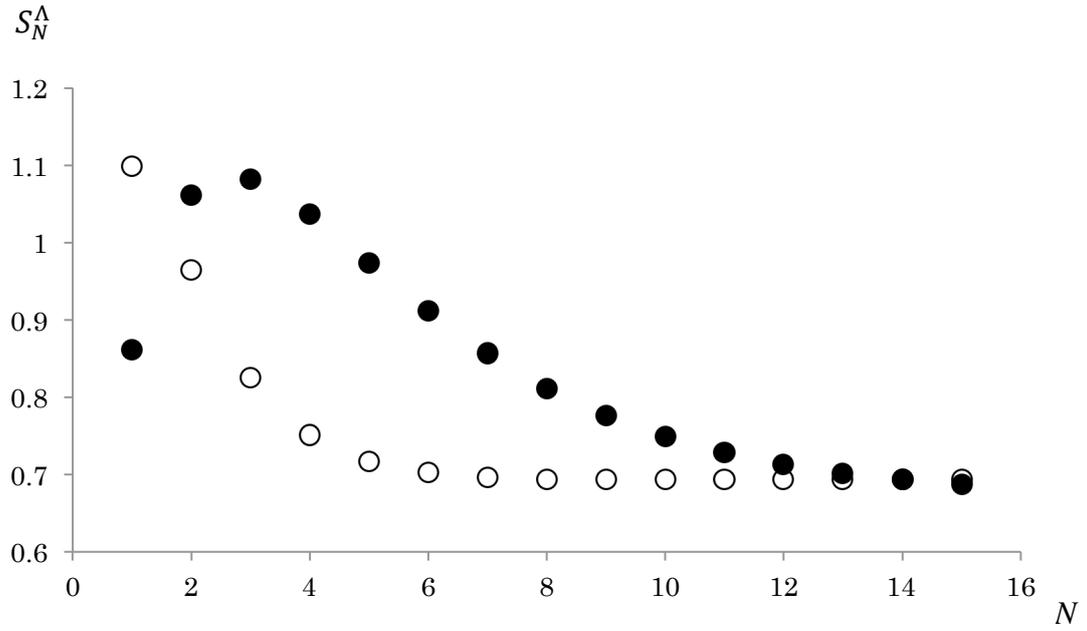

**Fig. 2.** Plots of $S_N^\Lambda$ in Eq. (44) with respect to $N$. ● and ○ correspond to the distributions (i) and (ii) mentioned in the text, respectively. All quantities are dimensionless.

# Corrigendum

We would like to eliminate the sentence "It is however know [2,3] ... in quantum theory." in the first paragraph of Section 1 on page 2, since this statement is misleading about positive semidefiniteness of a reduced density matrix.

We would like to thank Dr. V. Ambegaokar for pointing out this issue.